\documentclass[12pt]{article}

\usepackage[explicit]{titlesec}
\usepackage{etoolbox}
\usepackage{multicol}
\usepackage{hyperref}
\usepackage[natbibapa, nosectionbib]{apacite} 
\usepackage[symbol]{footmisc}
\usepackage{lmodern}
\usepackage[frozencache=true,cachedir=.]{minted} 
\usepackage[a4paper, portrait, margin=25mm]{geometry}
\usepackage{ragged2e}
\usepackage{comment}
\usepackage{adjustbox}
\usepackage{microtype}

\makeatletter
\patchcmd{\@maketitle}{\LARGE}{\Huge}{\typeout{OK 1}}{\typeout{Failed 1}}
\patchcmd{\@maketitle}{\large \lineskip}{\Large \lineskip}{\typeout{OK 2}}{\typeout{Failed 2}}
\makeatother

\makeatletter
\newcommand{\verbatimfont}[1]{\renewcommand{\verbatim@font}{\ttfamily#1}}
\makeatother

\titleformat{name=\paragraph, numberless}{}{\Large\normalfont\sffamily #1}{.5em}{}

\titleformat{name=\section}{}{\vspace{.8ex} \Large\normalfont\sffamily \thesection \hspace{0.2em} #1 }{.5em}{}
\titleformat{name=\subsection}{}{\vspace{.8ex} \large\normalfont\sffamily \thesubsection \hspace{0.2em} #1 }{.5em}{}

\newcommand{\newAuthor}[3]{
\noindent
#1 \\
#2 \\
#3 \\
}

\title{\textbf{\uppercase{ \justifying Partitura: A Python Package for Symbolic Music Processing}}}

\usepackage{xcolor}

\usepackage{xspace}
\usepackage{listings}

\def\ie{\textit{i.e.,}\xspace}
\def\eg{\textit{e.g.,}\xspace}

\def\MEI{\textsf{MEI}\xspace}
\def\Kern{\textsf{Humdrum **kern}\xspace}

\def\MIDI{\textsf{MIDI}\xspace}
\def\MusicXML{\textsf{MusicXML}\xspace}
\def\Music21{\textsf{Music21}\xspace}

\def\Partitura{\textsf{Partitura}\xspace}

\date{}

\usepackage{graphicx}
\graphicspath{{images/}}

\begin{document}

\maketitle

   
\begin{multicols}{3}
    \newAuthor{Carlos Cancino-Chac\'on\footnote{equal contribution.}}{Johannes Kepler University}{carlos\_eduardo.\\ cancino\_chacon@jku.at}
    
    \vspace{0.5cm}
    
    \newAuthor{Francesco Foscarin\footnotemark[1]}{Johannes Kepler University}{francesco.foscarin@jku.at}  
    
    \columnbreak
    
    \newAuthor{Silvan David Peter\footnotemark[1]}{Johannes Kepler University}{silvan.peter@jku.at}
    
    \vspace{1cm}
    
    \newAuthor{Maarten Grachten}{Independent Researcher}{maarten.grachten@gmail.com}     
    
    \columnbreak
    
    \newAuthor{Emmanouil Karystinaios\footnotemark[1]}{Johannes Kepler University}{emmanouil.karystinaios@jku.at}
    
    \vspace{1cm}
    
    \newAuthor{Gerhard Widmer}{Johannes Kepler University}{gerhard.widmer@jku.at}    
\end{multicols}

\paragraph{Abstract}

\emph{Partitura} is a lightweight Python package for handling symbolic musical information.
It provides easy access to features commonly used in music information retrieval tasks, like note arrays (lists of timed pitched events) and 2D piano roll matrices, as well as other score elements such as time and key signatures, performance directives, and repeat structures.
Partitura can load musical scores (in \MEI, \MusicXML, \Kern, and \MIDI formats),
 \MIDI performances, and score-to-performance alignments.
The package includes some tools for music analysis, such as automatic pitch spelling, key signature identification, and voice separation.
Partitura is an open-source project and is available at
\url{https://github.com/CPJKU/partitura/}.

\renewcommand{\thefootnote}{\arabic{footnote}}
\setcounter{footnote}{0}

\paragraph{Introduction}


%
In the past few years, symbolic music processing has been gaining increasing attention in the Music Information Research (MIR) community, with several music datasets of symbolic formats recently released, e.g. \citep{asap-dataset, micchi2020not, kong2020giantmidi}. 
Systems that target symbolic data are usually more efficient and easier to interpret than systems that target lower-level representation of music, such as audio files. This is not surprising, as sequences of notes are more compact and interpretable than sequences of amplitudes over time.
Symbolic formats can encode much more than a sequential note representation. Symbolically encoded musical scores arrange those notes in temporal and organizational structures such as measures, beats, parts, and voices. They can also explicitly represent dynamics and temporal directives and other high-level musical features such as time signature, pitch spelling, and key signatures.

While this rich set of musical elements adds useful information that can be leveraged by MIR systems, it also drastically increases the complexity of encoding and processing symbolic musical formats. 
Common formats for storage such as \MEI, \MusicXML, \Kern and \MIDI are not ideally suited to be directly used as input in MIR tasks\footnote{Though a few works exist that target directly \Kern files, \eg~\citep{Roman2019}, but some preprocessing is always required.}. 
Therefore, the typical data processing pipeline starts with parsing the relevant information from those files and putting it into a convenient data structure (\eg numerical arrays that can be used directly as input for machine learning or signal processing methods). 
Both operations require musical knowledge and can be very time-consuming, thus constituting a major barrier, especially for data-driven approaches that require a large dataset to be trained, and for researchers with limited musical background.

These problems have motivated us to develop \Partitura. Our goal is to simplify, as much as possible, all steps from the symbolic encoding to a convenient input data structure for a MIR system. \Partitura can straightforwardly produce standard data structures while still handling a complete set of symbolic music elements to create a customized one. 
\Partitura can parse symbolic representations of musical scores and performances from multiple file encodings (\MEI, \MusicXML, \Kern, and \MIDI) into Python objects to easily access their content. Moreover, it can produce commonly used data structures such as piano rolls and note arrays at different time resolutions.

The rest of this paper is structured as follows:
Section \ref{sec:related_work} highlights the differences between \Partitura and other python packages for processing symbolic musical formats. The package functionalities are detailed in Section~\ref{sec:partitura}, and in Section \ref{sec:getting_started} we provide a short usage example. 
Finally, in Section~\ref{sec:conclusions_and_future_work} we draw some conclusions on this paper and discuss possible future work.

\section{Related Work}\label{sec:related_work}

Among the available Python packages for parsing and processing music in symbolic formats, there are two that stand out in terms of popularity and usability, \emph{Pretty MIDI} and \emph{music21}.

\emph{Pretty MIDI}~\citep{raffel2014intuitive} is a Python package that focuses on the analysis, modification, and generation of MIDI data in a fast and straightforward way.
A strong feature of PrettyMIDI is its ability to easily extract MIDI properties such as the position of beats and downbeats, key and time signatures, and to produce piano roll representation with a specific sample frequency.
\Partitura follows the PrettyMIDI philosophy of speed and simplicity
but extends it to other symbolic formats of musical scores, and to the other notation elements they contain. Moreover, while PrettyMIDI only represents time in seconds, \Partitura can work with other time units such as beats and quarter notes.

Another well-known python package for handling both MIDI and richer symbolic encodings of musical scores is \emph{music21}
~\citep{music21}. Indeed, music21 has been developed and supported for many years now. It offers a robust parser for many file formats, and support for many ``advanced'' score elements such as nested tuples and beamings. Among the package goals, there are advanced modifications of musical scores, such as transpositions, pitch respelling, insertion, and deletion of voices and measures. All this is supported by an internal representation based on nested containers called Streams that model the hierarchical temporal and organizational structure of the score in measures, voices, and parts. \Partitura does not aim at rebuild such a complete and complex framework, instead, it focuses on  a different goal: lightweight extraction of features that are relevant for MIR research, typically sequential representations of score elements such as piano rolls or note arrays. To efficiently target this objective, \Partitura uses a much simpler, but nonetheless complete, sequential representation of musical scores, with musical elements arranged in a timeline.

\section{Partitura}\label{sec:partitura}

\Partitura can handle three symbolic data types: musical scores, performances, and score-to-performance alignments. The score contains a representation of music, highly structured in staves, measures, beats, and voices, and express durations in \textit{musical units} quantized to fractions of quarter notes and beats.
The performance is a sequential representation of musical events expressed on a continuous timeline and not quantized to fixed values. Alignments between the two formats can be done at note- or time-level (\eg beat and measure). 
Different file formats can be parsed into dedicated internal representations to offer easy access to the file content. A set of functions creates data structures that are often used in MIR research. Finally, Partitura offers some music analysis tools.

\subsection{Internal Data Structures}\label{sec:internal_data_structure}

\begin{figure}
    \centering
    \includegraphics[width=0.86\textwidth]{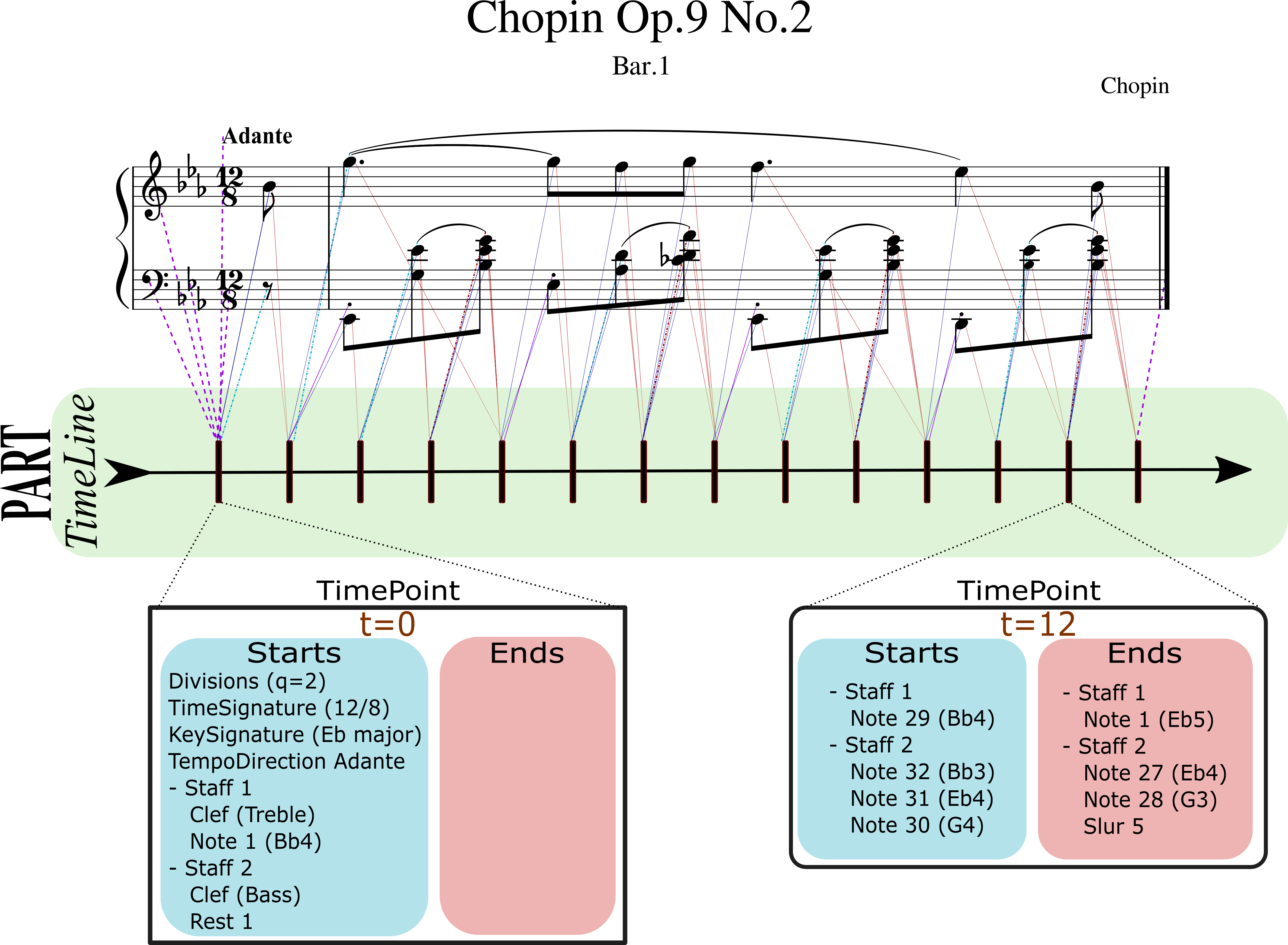}
    \caption{Schematic representation of a Part. The blue lines represent the starting times of the objects in the score and the red lines represent the end times.}
    \label{fig:timeline_example}
\end{figure}

Different internal data types are built to represent  scores, performances, and score-to-performance alignments.

For a score, Partitura uses three main classes: \textit{TimePoints}, \textit{TimedObjects}, and \textit{Parts}.
At the highest level, there are one or more \textit{Part} objects, possibly grouped by \emph{PartGroup} objects. 
\textit{Parts} are typically associated with instruments, and each \textit{Part} may have one or more staves.
Each \textit{Part} contains a timeline that encapsulates a sequence of \textit{TimePoint} objects, each denoting a temporal position in the score.
\textit{TimePoints} encode score time in non-negative integer units.
The relation of this unit to a quarter note is chosen
such that any temporal position present in the score can be represented in integer values.

Musical elements (for example, a \emph{Note}) are added to the timeline by registering them with the \textit{TimePoints} corresponding to their start and end positions.
Any element registered with two \textit{TimePoints} is a \textit{TimedObject}.
\Partitura represents a large set of score elements as subclasses of the \textit{TimedObject}, \eg notes, rests, time signatures, key signatures, slurs, measures, tempo and loudness directives. Figure \ref{fig:timeline_example} shows a schematic representation of a \textit{Part} object and its components.


In contrast to scores, the performance is inherently sequential and can be represented in a simpler structure. 
\Partitura uses \textit{PerformedPart} objects that consist of two ordered containers which store notes and MIDI control information. 
A note object of a \textit{PerformedPart} is a dictionary encoding  MIDI note parameters (onset, offset, velocity, pitch, channel, and track) as well as a deterministically generated unique note identifier.

Score-to-performance alignments are represented with a \textit{Part}, a \textit{PerformedPart}, and a sequence of alignment pairs.
Each alignment pair encodes a link between a note ID (or time position) in the score and a note ID (or time position) in the performance.

\subsection{Supported File Formats}\label{sec:supported_formats}

\Partitura can parse score formats such as \MEI, \MusicXML, \Kern and produce \textit{Part} objects. 
The case of \MIDI files is more complex, as they can encode either a performance or a bare-bones scores representation \citep{midi_spec}.
\Partitura loads MIDI scores and MIDI performances into \textit{Part} and \textit{PerformedPart} objects respectively.
As far as output file formats are concerned, Partitura can produce \MusicXML and MIDI files from \textit{Parts} and MIDI files from \textit{PerformedParts}.

\Partitura supports import and export functionality for match files, a format for encoding symbolic score-to-performance music alignments~\citep{matchpaper}.
Furthermore, \Partitura parses simpler alignment file formats such as the .match and .corresp files proposed by \citet{nakamura2017performance}.

\subsection{Generated Data Structures}\label{sec:generated_data_structures}
Although convenient for lossless representation of score time, the internal representation of time points and durations as integers is not particularly meaningful from a musical perspective. 
For this reason, \Partitura can output temporal positions and durations in two other units: quarter notes and beats.
For example, the upper-staff notes of the score in Figure~\ref{fig:timeline_example} would have a temporal position of $[0,1,5,6,7, \dots]$ if we are considering quarter notes, $[-1,0,4,5,6 \dots]$ if we are considering ``slow-tempo beats'' (12 beats for the measure) or $[-0.333, 0, 1, 0.333, 0.666, \dots]$ if we are considering ``fast-tempo beats'' (4 beats for the measure).
Mappings between various time units are readily available as \textit{Part} methods. 

\begin{figure}
    \centering
     \begin{minipage}{\textwidth}
        \begin{multicols}{2}
            \begin{adjustbox}{width=\columnwidth, center}
            \begin{tabular}{l|c c c c c c}
            \textbf{Instances} & \textbf{Id} & \textbf{Onset} & \textbf{Duration} & \textbf{Pitch} & \textbf{Time Signature} & \dots\\
            $note_1$ & $id_1$ & $on_1$ & $dur_1$ & $pi_1$ & $ts_1$ &  \\ 
            $note_2$ & $id_2$ & $on_2$ & $dur_2$ & $pi_2$ & $ts_2$ &  \\
            $\vdots$ & $\vdots$ & $\vdots$ & $\vdots $ & $\vdots$ & $\vdots$  &  \\
            $note_N$ & $id_N$ & $on_N$ & $dur_N$ & $pi_N$ & $ts_N$ &  \\
            $\vdots$ & $\vdots  $ & $\vdots $ & $\vdots$ & $\vdots$ & $\vdots$ & \\
        \end{tabular}
        \end{adjustbox}
        
        \columnbreak
        
        \includegraphics[width=\columnwidth]{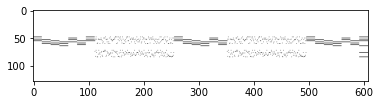}
            
        \end{multicols}
    \end{minipage}
    \caption{An abstract example of a note array (left) and a piano roll(right).}
    \label{fig:note_array}
\end{figure}

\Partitura can automatically generate two data structures that are commonly used in MIR tasks: \textit{note arrays} and \textit{piano rolls matrices} (see Figure~\ref{fig:note_array}). 
A note array is an ordered sequence of note features. Such features can include note descriptors (\eg midi-pitch, pitch-spelling, onset position, voice, and duration), but  also context information like metrical position, time signature, and key signature. Users can choose among these features the ones that are related to their application.
A use case is demonstrated in Figure~\ref{fig:import_mei}
From this representation, we can build a dedicated word encoding of the musical score, as done by \cite{hawthorne2018maestro}.
A piano roll is a matrix of shape $(\textrm{number of pitches} \times \textrm{number of time frames} )$ where the length of a time frame can be set as fractions of beats or quarter notes. 
For example, if we consider 4 frames per quarter note, and the piano range, the score of Figure~\ref{fig:timeline_example} would produce a piano roll of shape $(88 \times 28)$.  
This representation is widely used in the MIR community, for example, by \cite{bach_doodle}.
Built-in methods to create \textit{note arrays} and \textit{piano rolls matrices} are available for both \textit{Part} and \textit{PerformedPart} objects.
For efficient processing in Python, note arrays and score piano rolls are numpy arrays.\footnote{\url{https://numpy.org/}}

\subsection{Music Analysis and Repetition Unfolding Tools}\label{sec:musica_analysis_and_repetition_unfolding_tools}

\Partitura includes some tools for music analysis that are intended to fill in missing information with plausible values, for instance, when loading a score from a MIDI file. The list of available tools includes
the Krumhansl--Shepard algorithm~\citep{krumhansl1990} for key signature estimation, the \emph{ps13s1} algorithm~\citep{Meredith:ps13} for pitch spelling, and \emph{VoSA}~\citep{Chew2004} for polyphonic voice estimation.
To our knowledge, this is the first publicly available Python implementation of ps13s1 and VoSA.

Musical scores often encode repetition structures with repeat signs, Volta brackets, and navigation directions such as al Coda, dal Segno, da Capo, or al Fine. On the other side, music performances are ``unfolded'' and multiple possible unfoldings can exist for a piece, as players often decide to skip some repetitions.
\Partitura supports the generation of such unfoldings from a score's repetition structure
and their conversion into a new \textit{Part} object.

\section{Getting Started}\label{sec:getting_started}

In this section, we present a quick introduction to the usage of the Partitura package.
For more examples of use cases, as well as a more detailed description of the elements of the package, please refer to the online documentation.\footnote{\url{https://partitura.readthedocs.io/}}
A hands-on tutorial can be found on Google Colab.\footnote{\url{https://tinyurl.com/partituratutorial}}

The Partitura package can be installed from Python Package Index\footnote{\url{https://pypi.org}} with the command \verb|<pip install partitura>| or directly from the source code available on \href{https://github.com/CPJKU/partitura/}{Github}.

\subsection{Importing Files}

\begin{figure}[t]
\noindent
\centering
\begin{minipage}{.8\linewidth}
\begin{minted}[
frame=lines,
framesep=2mm,
baselinestretch=1.2,
fontsize=\small,
linenos
]{python}
import partitura as pt

# Load score
score = pt.load_score("chopin_op9_no2.mei")
# Load MIDI as a performance
performance = pt.load_performance("chopin_op9_no2_perf.mid")
# Load Alignment
performance, alignment = pt.load_match('chopin_op9_no2.match')
\end{minted}
\end{minipage}
\caption{Importing Files.}
\label{fig:import_export}
\end{figure}

As mentioned in Section \ref{sec:internal_data_structure}, \Partitura treats musical scores and performances differently, and this is reflected in how scores and performances are imported.
\Partitura includes a generic \verb|load_score| method for loading files as scores (\ie as \textit{Parts}), as well as a \verb|load_performance| method for loading files as performances (\ie as \textit{PerformedParts}).
These generic methods infer the format of the input file automatically. 
Additionally, there are individual methods for loading supported formats (\eg \verb|load_musicxml|, \verb|load_mei|, \verb|load_kern| for \MusicXML, \MEI, \Kern files, respectively).
For \MIDI files, Partitura provides both \verb|load_score_midi| and \verb|load_performance_midi| methods.
By doing so, we expect users to know what kind of information they would like to extract from MIDI files.
As mentioned in Section \ref{sec:musica_analysis_and_repetition_unfolding_tools}, we use the included music analysis tools to infer plausible values for the missing information (especially pitch spelling and voice information) in MIDI files imported as scores.
Figure \ref{fig:import_export} shows an example of loading files.

\subsection{Computing Note Arrays and Piano Rolls}

\begin{figure}[t]
\noindent
\centering
\begin{minipage}{.8\linewidth}
\begin{minted}[
frame=lines,
framesep=2mm,
baselinestretch=1.2,
fontsize=\small,
linenos
]{python}

import partitura as pt
import matplotlib.pyplot as plt

# Import score
part = pt.load_score("chopin_op9_no2.mei")
# Get note array
note_array = part.note_array(include_time_signature=True)
# get piano roll
pianoroll = pt.utils.compute_pianoroll(part, 
                                       time_div=8, 
                                       piano_range=True)
# plot piano roll
plt.spy(pianoroll)
\end{minted}

\textbf{Output: }

\begin{center}
    \includegraphics[width=\textwidth]{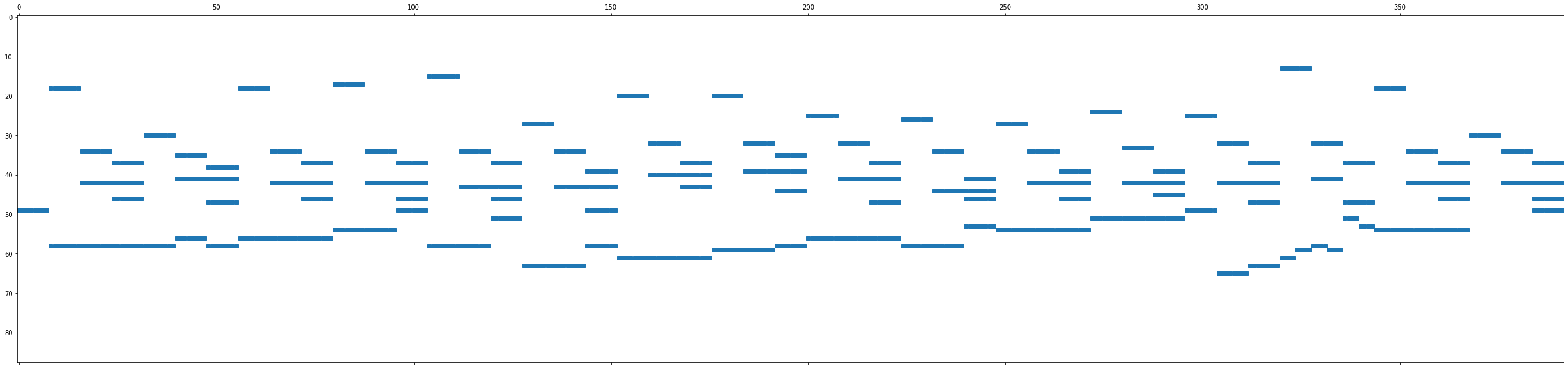}
\end{center}
\end{minipage}
\caption{Extracting note arrays and piano rolls (score of Figure~\ref{fig:timeline_example}).}
\label{fig:import_mei}
\end{figure}

Figure \ref{fig:import_mei} shows an example of extracting note arrays and piano rolls from a score (the syntax is the same for performances).
This example illustrates the philosophy of Partitura of reducing the most common operations on symbolic music to one-line Python commands.

In Partitura, note arrays are implemented using Numpy structured arrays,\footnote{\url{https://numpy.org/doc/stable/user/basics.rec.html}} arrays in which each column can have a different datatype.
By default, note arrays generated from scores include onset and duration information (in beats, quarters and divs), MIDI pitch, voice and note ID. 
Note arrays generated from performances include onset and duration in seconds, MIDI pitch, velocity track and channel; and Note ID.
The \verb|note_array| method in \textit{Parts} and \textit{PerformedParts} can receive a list of user defined callable methods which can compute other features at the note level (\eg scale degree, etc.).

Since computing piano rolls results in very sparse matrices, \Partitura computes piano rolls as Scipy sparse matrices.\footnote{\url{https://docs.scipy.org/doc/scipy/reference/generated/scipy.sparse.csr_matrix.html}}
The method \verb|compute_pianoroll| can also specify the desired resolution of the piano roll by specifying the number of sub-divisions for each time unit with the \verb|time_div| argument. 

\subsection{Music Analysis Tools}

As mentioned in Section \ref{sec:musica_analysis_and_repetition_unfolding_tools}, \Partitura includes tools for estimating key signature, pitch spelling, voice information, and tonal tension, which can be found in the \verb|partitura.musicanalysis| module. 
The methods in this module accept \textit{Parts},  \textit{PerformedParts} or note arrays as input and return a Numpy structured array with the estimated information, except for \verb|estimate_key| which returns the estimated key signature as a string (\eg \verb|'Cm'|, \verb|'F#m'|, etc.).
Figure \ref{fig:music_analysis} shows an example of how to compute this information from a score.

\begin{figure}[t]
\noindent
\centering
\begin{minipage}{.8\linewidth}
\begin{minted}[
frame=lines,
framesep=2mm,
baselinestretch=1.2,
fontsize=\small,
linenos
]{python}

import partitura as pt
# Load score
part = pt.load_score("chopin_op9_no2.mei")
# Estimate key signature
key_name = pt.musicanalysis.estimate_key(part)
# Estimate pitch spelling
pitch_spelling = pt.musicanalysis.estimate_spelling(part)
# Estimate voice information
voices = pt.musicanalysis.estimate_voices(part)
# Compute tonal tension
tonal_tension = pt.musicanalysis.estimate_tonaltension(part)
\end{minted}
\end{minipage}
\caption{Music analysis tools in Partitura.}
\label{fig:music_analysis}
\end{figure}

\section{Conclusions and Future Work}\label{sec:conclusions_and_future_work}

In this paper, we presented \Partitura, a Python package for handling symbolic music information that requires minimal music expertise.
This package can parse common symbolic music formats, like \MusicXML, \MEI, and \MIDI, and conveniently represent them as Python objects that are easy to manipulate in automatic data pipelines. Moreover, it can straightforwardly produce the most commonly used data structures for MIR tasks.
To the best of our knowledge, Partitura is the only Python library that can handle alignments between scores and corresponding performances.

Future work will be in the direction of making Partitura file parsers more robust to bad encoding practices that are unfortunately very frequent in symbolic musical scores. Moreover, support for more score elements and different data structures will be added to keep Partitura on track with new needs and demands from the research community.
We are working on making Partitura faster and more efficient by optimizing existing methods and including support for parallel data processing.
Another functionality that will be added is the automatic score unfolding to match the repetition structure of a given performance. Finally, we will develop more analysis tools to infer high-level score elements, with the final goal of being able to ``scorify'' a performance or an incomplete score representation.

\paragraph{Acknowledgements}
This project receives funding from the European Research Council (ERC) under the European Union's Horizon 2020 research and innovation programme, grant agreement No 101019375 (\textit{Whither Music?}).


\paragraph{References}

\bibliographystyle{apacite}
\bibliography{biblio}

\end{document}